# The Evaluation of Mechanical Properties of LB650 Cavities*

J. Holzbauer, G. Wu[†], H. Park, K. McGee, A. Wixson, T. Khabiboulline, G. Romanov, S. Adams,
D. Bice, S. K. Chandrasekaran, J. Ozelis, I. Gonin, C. Narug, R. Thiede, R. Treece, C. Grimm,
Fermilab, Batavia, USA

*Abstract*

The PIP-II project's LB650 cavities could potentially be vulnerable to mechanical deformation because of the geometric shape of the cavity due to reduced beta. The mechanical property of the niobium half-cell was measured following various heat treatments. The 5-cell cavities were tested in a controlled drop test fashion and the real-world road test. The result showed that the 900 °C heat treatment was compatible with cavity handling and transportation during production. The test provides the bases of the transportation specification and shipping container design guidelines.

## INTRODUCTION

The PIP-II linac consists of several types of cryomodules made of HWR, SSR1, SSR2, LB650, and HB650 cavities [1]. The dynamic heat load of the cavities in cryomodules is primarily of SSR2, LB650, and HB650 cryomodules, as shown in Figure 1 [2].

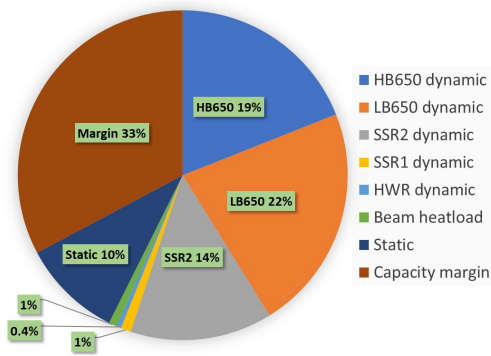

Figure 1: PIP-II linac total heat load includes the static, dynamic, beam, and static heat loads.

The 650 MHz cavities have high $Q_0$ processing procedures to minimize the total heat load. The high $Q_0$ processing includes nitrogen doping or medium-temperature furnace baking (Mid-T furnace baking). As a result, cavities receiving the high $Q_0$ processing have a higher surface resistance sensitivity to the trapped magnetic field [3], as seen in Table 1.

It has been demonstrated that a successful magnetic shielding design and magnetic hygiene implementation can reduce the residual magnetic field seen by cavities in a cryomodule to ~ 1.5 mG [4].

However, thermoelectric current could flow through cavities during the cryomodule cooldown, generating up to 8 mG residual magnetic fields [5]. If trapped during the superconducting transition, those residual magnetic fields could dramatically increase the cavity dynamic heat load regardless of the processing method.

Table 1: Surface resistance sensitivity of magnetic field trapping

| Processing | Surface Resistance Sensitivity [nΩ/mG] |
|---|---|
| EP only | 0.4 |
| N-doped | 1.4 |
| Mid-T furnace baking (350 °C) | 1.0 |

The fast cooldown has been demonstrated in LCLS-II cryomodules that can effectively expel the thermoelectric current-induced magnetic field [6]. In addition, the heat treatment of niobium cavities at 900 °C for 3 hours has been confirmed for several LB650 cavities that can expel much of the residual magnetic field, while the cavities treated at 800 °C for 3 hours resulted in the field mainly being trapped [7].

The geometry of the PIP-II projects' 650 MHz cavities, particularly the LB650 cavities, indicated that the cavity's effective mechanical strength might be weaker than that of ILC 1.3 GHz cavities. A 900 °C treatment may decrease the niobium yield strength to a level that could deform the cavity during handling, transportation, and pressure testing. Therefore, a validation plan was developed to verify the cavity's mechanical strength. The plan included the following tests:

- Sample mechanical tests
- Transportation tests on local highways
- Controlled drop tests
- Pressure tests of a jacketed cavity.

## EXPERIMENT

### Sample preparation

Niobium samples were made from the same niobium sheets used to fabricate LB650 Cavities. They were the corner material during the cavity sheet preparation. The material has a 4.5 mm thickness. The samples were prepared as full-size dog bones according to the ASTM standard. The samples were chemically etched as with cavities, followed by hydrogen de-gas heat treatment at either 800 °C or 900 °C for three hours.

___
* Work supported by Fermi Research Alliance, LLC, under Contract No. DE-AC02-07CH11359 with the U.S. Department of Energy, Office of Science, Office of High Energy Physics
† genfa@fnal.gov

## Transportation tests

Bare cavities were used in the transportation test. The simulation showed that peak stress value and location are very similar between a jacketed cavity and a bare cavity supported only with the end supports of the cavity support cage, as shown in Figure 2.

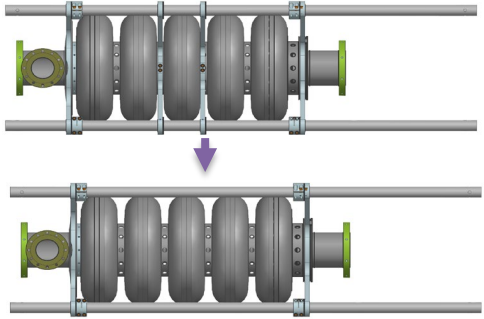

Figure 2: Cavity support cage with the two end supports.

Cavities were pre-stretched to match the vacuum load. They sat in containers fitted with polyethylene foam with three 3-D accelerometers attached to the cavities. Cavity passband frequency and field flatness were measured before and after the road tests. Two different area highway routes were chosen. Each round trip took about two hours in a flatbed box truck.

## Controlled drop tests

Bare cavities, again, were used in the controlled drop test. The simulation showed that the peak stress is located at the coupler end, as shown in Figure 3.

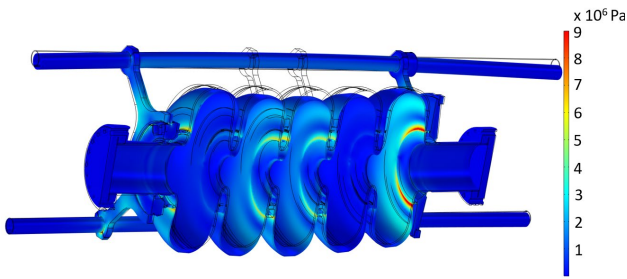

Figure 3: A stress analysis with the cavity in a support cage with the two end supports. The load was at 1 g.

A controlled drop test was developed to drop the coupler side of the cavity while the cavity was in a transportation container, the same as in the road tests. As shown in Figure 4, one side of the transportation container is lifted to a certain height. A quick release allows the container to fall to the ground, creating a shock like in a road condition. The height can be controlled to adjust the shocks experienced by the cavity. The drop test can be repeated as needed. The shock was increased as the height was increased. When the shocks reached 5 g, the height was maintained 10 times. After the repeated 5 g shocks, the height was increased until the cavity experienced cell deformation or the shock reached the plateau when the increased height no longer resulted in an increased shock magnitude.

Cavity passband frequencies were monitored in situ to assess the potential cavity deformation. From previous experience, the cavity passband frequency spread is a good indicator of field flatness deviation. A 10 kHz passband frequency deviation would be measurable for a field flatness change which indicates a potential cavity cell deformation. Cavity field flatness is measured before and after the controlled drop tests.

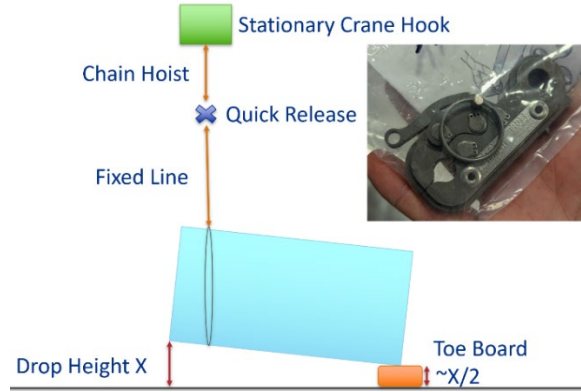

Figure 4: Drop test configuration.

## Pressure tests of a jacketed cavity

After a cavity is jacketed, pressure tests are a standard quality control step during cavity fabrication. The pressure test applies to the helium space between the cavity and the helium vessel. The pressure test includes two "round trip cycles." Each cycle has pressure increasing from zero to 35 psi and decreasing to zero. Cavity passband frequencies are monitored during the pressure test.

# RESULTS AND DISCUSSION

## Sample mechanical properties

A total of 15 samples were tested. Their results are shown in Table 2. Both 800 °C and 900 °C treated samples show reduced niobium yield strength. Yet, there was no difference in the average yield strength between the two temperatures. All values exceeded the required minimum yield strength of 38 MPa.

Table 2: Niobium sample yield strength

| As-received Samples | 800 °C treated samples | 900 °C treated samples |
|---|---|---|
| 82.0 | 49.6 | 40.7 |
| 87.6 | 53.8 | 49.0 |
| 85.5 | 55.2 | 69.6 |
| 78.6 | 49.0 | 47.6 |
| 91.7 | 43.4 | 44.1 |
| 85.1 | 50.2 | 50.2 |

The last row indicates an average value.

## Transportation Tests

Three road tests were conducted for a 900 °C treated cavity. Route 1 was selected due to significant construction activities with very unfavorable road conditions. Route 2 showed normal road conditions.

Table 3 shows the cavity was safe in a standard transportation container with polyethylene foam material.

Table 3: Field flatness results of the road tests

| Trip | Road Condition | Field Flatness before | Field Flatness after |
|---|---|---|---|
| Route 1 | Multiple 5g | 92% | 98% |
| Route 1 | Single 10 g | 98% | 96% |
| Route 2 | Single 5 g | 96% | 95% |

## Controlled Drop Tests

During the controlled drop tests, an acceptance criterion was set to field flatness >90% and frequency spread < 10 kHz.

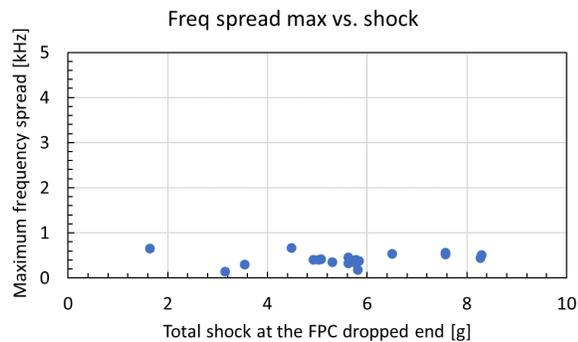

Figure 5: A LB650 cavity drop test with shocks reached > 8 g.

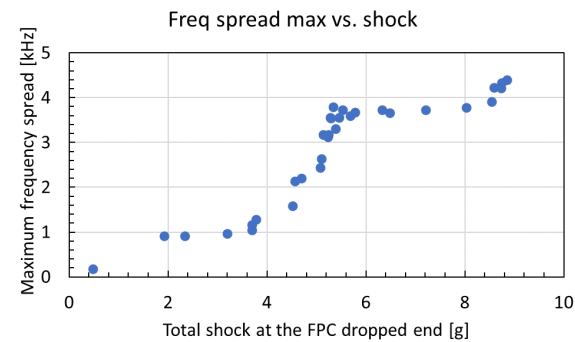

Figure 6: A LB650 cavity drop test with shocks reached 9 g.

Several cavities treated at 900 °C temperatures were drop tested. All have their field flatness > 90%. Their maximum passband frequency spreads are all within 10 kHz. Three cavities experienced <3 kHz, a negligible amount indicating there was no detectable cavity deformation with repeated 5 g shocks and up to a 10 g shock. Figure 5 shows a 900 °C treated LB650 cavity that experienced no detectable cell deformation up to 8.2 g. Figure 6 shows a 900 °C treated LB650 cavity that experienced no detectable cell deformation up to 8 g. For that cavity, when the shock was increased to above 8 g, the frequency spread started to increase, indicating that cell deformation started

## Jacketed Cavity Pressure Tests

A jacketed LB650 cavity was pressure tested. Figure 7 shows the two cycles of the pressure test trip. The fundamental mode frequency change was < 5 kHz, meeting the requirement of < 10 kHz. The field flatness changed from 92% to 94%, which met the >90% requirement and was within the measurement error.

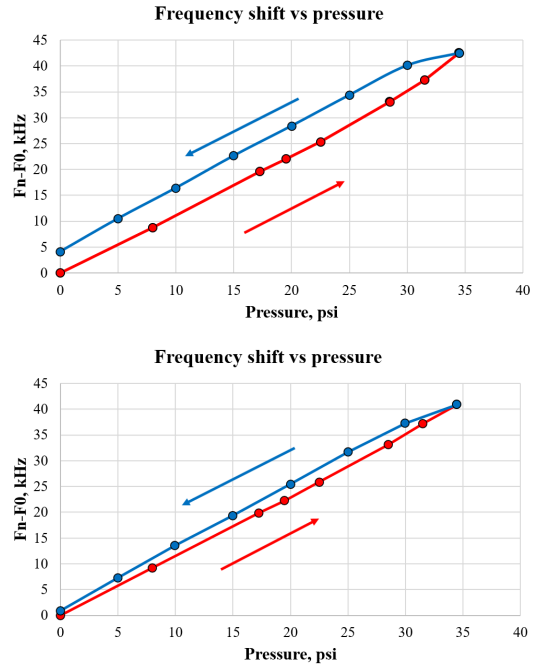

Figure 7: Fundamental mode frequencies measured during the pressure tests. The upper figure is cycle 1, and the lower figure is cycle 2. The arrows indicate the measurement sequence.

## CONCLUSION

A series of tests, including samples, transportation, controlled shocking, and jacketed cavity pressure tests, indicated that the 900 °C three-hour heat treatment of LB650 MHz bare cavities is acceptable for jacketing, cavity handling, and transportation.

Cavities would experience no detectable cell deformation up to 8 g.

The result further provided validation of the transportation specification of cavities and cryomodules.

## ACKNOWLEDGMENTS

This manuscript has been authored by Fermi Research Alliance, LLC, under Contract No. DE-AC02-07CH11359 with the U.S. Department of Energy, Office of Science, Office of High Energy Physics.